\documentclass[11pt,showkeys,showpacs,a4paper,aps,amssymb,floatfix]{revtex4-1} 
\usepackage{amsmath}
\usepackage{amsfonts}
\usepackage{textcomp}
\usepackage{graphicx}
\usepackage{epstopdf}
\usepackage{algorithm}
\usepackage{algpseudocode}
\usepackage[english]{babel} 
\usepackage[latin1]{inputenc}
\usepackage{hyperref} 
\usepackage{natbib}  
\usepackage[usenames]{color} 
\usepackage{float}
\usepackage{wrapfig}

\let\originaleqref=\eqref
\renewcommand{\eqref}{equation~\originaleqref}
\scrollmode

\graphicspath{{./../Figures-2D/}}
\pdfoutput=1
\begin{document}

\title{Plasmonic and photonic crystal applications of a vector solver for 2D Kerr nonlinear waveguides}
\author{Mahmoud M. R. Elsawy}
\author{Gilles Renversez$^*$}
\affiliation{Aix Marseille Univ, CNRS, Centrale Marseille, Institut Fresnel, 13013 Marseille, France}
\email[]{gilles.renversez@univ-amu.fr}

\date{\today}
 
\begin{abstract}
We use our vector Maxwell's nonlinear eigenmode solver to study  the stationary solutions in 2D cross-section waveguides with Kerr nonlinear cores. This solver is based on the fixed power algorithm within the finite element method.  First,  studying nonlinear plasmonic slot waveguides, we demonstrate that, even in the low power regime, 1D studies may not provide accurate and  meaningfull results compared to 2D ones. Second, we study at high powers the link between the nonlinear parameter $\gamma_{nl}$ and the change of the nonlinear propagation constant $\Delta \beta$. Third,  for a specific type of photonic crystal fiber, we show that a non-trivial interplay between the band-gap edge   and the nonlinearity takes place. 
\end{abstract}


\maketitle
\thispagestyle{plain}
	
Merging plasmonics  or photonic crystals with nonlinear structures has been predicted to be a promising choice in designing compact nonlinear optical devices~\citep{dekker2007ultrafast,kauranen_nonlinear_2012}, due to the enhancement of the nonlinearity ensured by the achieved strong light confinement. Among the different nonlinear plasmonic configurations~\citep{walasik_stationary_2014}, the nonlinear plasmonic slot waveguide (NPSW) in which a nonlinear core of Kerr-type is surrounded by two metal claddings has received a great attention. Different numerical and analytical methods have been developed to provide complete and full studies for the one-dimensional (1D) NPSWs with isotropic~\citep{davoyan_nonlinear_2008,rukhlenko_exact_2011,salgueiro_complex_2014,walasik_plasmon-soliton_2016-1} and anisotropic nonlinear cores~\citep{elsawy_study_2017}. 

For the first time,  using our rigorous and accurate nonlinear solver, we provide the results directly based on the nonlinear modes in realistic 2D cross section NPSWs and compare them with those from a 1D nonlinear model~\cite{elsawy_study_2017}.
It is worth mentioning that the main nonlinear modes in all-dielectric 2D slot waveguide have been investigated using a nonlinear solver~\cite{fujisawa_guided_2006}, however, in the studied structure the field confinement decreases for high power.

Another illustration of the usefulness of our vector nonlinear solver is the study of a photonic crystal fiber (PCF) with a Kerr-type nonlinear matrix with a defect as fiber core. Similar structures have already been studied in the scalar case~\citep{ferrando_spatial_2003} or even in the vector case~\citep{fujisawa2003PCF-NL}. Nevertheless, due to the type of the studied PCF and of the used core defect, only expected results like a strong focusing at very high powers are obtained while in our case a non-trivial effect is obtained at realistic powers.
	
The paper is organized as follows, first, we briefly describe our rigorous 2D vector nonlinear solver to compute the stationary nonlinear solutions propagating in  2D waveguides with a Kerr-type nonlinear material.

Second, we demonstrate the importance of a 2D nonlinear solver, by showing that 1D modelling of NPSW is not enough to understand and quantify the nonlinear characteristics of realistic 2D configurations. Third, based on our nonlinear solver, we present a general interpretation of the Kerr-type nonlinear parameter $\gamma_{nl}$ as a function of the total power for two nonlinear plasmonic waveguides, and we show that the usual definitions usually used in the literature are valid only in limited power range. Fourth, we show that in a PCF with an ad-hoc defect as core fiber, a defocusing behaviour can be observed even if a focusing Kerr-nonlinearity is considered. 

\begin{figure}[bth!]
	\centering
	\includegraphics[width=\columnwidth, trim=0 0 0 0]{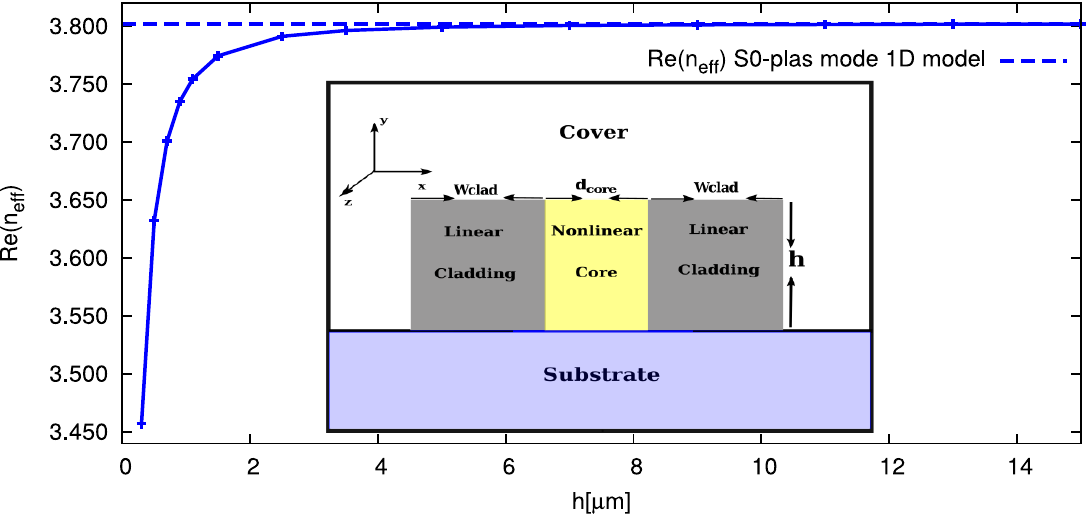}
	\caption{$\Re e(n_{eff})$  obtained from the linear 2D slot waveguide (see the inset) with linear metal claddings as a function of the waveguide height $h$ for the linear fundamental symmetric mode S0-plas. Here, $d_{core}=400$ nm (core thickness) and the cladding width $\text{W}_{\text{clad}}=300$ nm. See the text for the other parameters. The horizontal dashed line represents the results from the 1D model (invariant along $y$ direction).}
	\label{fig:structure2D}
\end{figure} 

In our model, we consider monochromatic waves propagating along the $z$ direction in 2D invariant cross section	waveguides such that all field components evolve proportionally to $\exp[i( \beta z - \omega t)]$ with 
$\omega$ is the angular frequency and $\beta$ denotes the complex propagation constant. We consider complex and diagonal nonlinear permittivity tensor $\tilde{\bar{\bar{\epsilon}}}=\tilde{\bar{\bar{\epsilon}}}_{t}+\tilde{\bar{\bar{\epsilon}}}_{z}$ such that:
\begin{multline}
  \label{eq:nl_perm_general}
  \tilde{\bar{\bar{\epsilon}}}_{t}=\left[\epsilon_{x}~~  \epsilon_{y} ~~ 0 \right]+i\left[\epsilon^{\prime \prime}_{x}~~\epsilon^{\prime \prime}_{y}~~0   \right],~~~ \tilde{\bar{\bar{\epsilon}}}_{z}=\left[0~~  0 ~~ \epsilon_{z} \right]+i\left[0~~0~~\epsilon^{\prime \prime}_{z}   \right]\\
  \text{with}~~{ {\epsilon_{j}}(x,y)}={{\epsilon_{jj}}(x,y)}+{ {\alpha_{jj}}(x,y)}  | {\mathbf{E}(x,y)} |^{2}~~~\forall j \in {x,y,z}.
\end{multline}  
Here, $\tilde{\bar{\bar{\epsilon}}}_{t}$ and $\tilde{\bar{\bar{\epsilon}}}_{z}$ represent the complex nonlinear permittivity tensors along the transverse and the longitudinal directions, respectively. $\epsilon_{jj}(x,y)$ is the linear part of the permittivity, ${ {\alpha_{jj}}(x,y)}$ is the corresponding nonlinear parameter~\citep{boyd_nonlinear_2008}, and $\mathbf{E}(x,y)$ is the total complex amplitude of electric field which can be also decomposed into transverse and longitudinal components $\mathbf{E}(x,y)=\mathbf{E}_{t}(x,y)+{E}_{z}(x,y)~\mathbf{\hat{z}}$ .
In order to compute the linear and nonlinear modes we use the finite element method (FEM) using the code we developed (based on gmsh/getdp free softwares~\cite{dular_general_1998}) to study several types of waveguides including nonlinear ones with Kerr regions like microstructured optical fibers~\citep{drouart_spatial_2008} (2D scalar model) and 1D nonlinear plasmonic waveguides with isotropic and anisotropic Kerr-type nonlinearities~\citep{walasik_stationary_2014,elsawy_study_2017}. We use the fixed power algorithm in order to treat the nonlinearity, in which the total power is an input and the outputs are the nonlinear propagation constant and the corresponding nonlinear field profile. Since we are dealing with a nonlinear problem, in which the superposition principle does not hold, therefore, in the frame of the fixed power algorithm, all the amplitudes of the electromagnetic field components must be scaled correctly using a scaling factor, taking into account the change in the effective index and the nonlinear field profiles~\citep{rahman_numerical_1990,ferrando_spatial_2003,elsawy_study_2017}. 
Therefore, we assume that the transverse and longitudinal components $\mathbf{E}_{t}$ and $E_{z}\mathbf{\hat{z}}$ are given by:
\begin{equation}
\label{eq:2D_functions_chi_Et_Ez}
	\mathbf{E}_{t}=\chi \pmb{\phi}_{t}~~~~~~\text{and}~~~
	E_{z}\mathbf{\hat{z}}=i\beta \left( \chi  \phi_{z} \right) \mathbf{\hat{z}}.
\end{equation}
where  $\pmb{\phi}_{t}$  together with $ \phi_{z} $ and $\beta$ are, repectively, the eigenvector and eigenvalue of an eigenvalue problem written in the frame of the FEM  (see Eqs.~(4.26) and (4.27) of the chapter~4 in Ref.~\citep{zolla_foundations_2012}). $\chi$ is a scaling factor that must be computed for a given input total power $P_{tot}$:
\begin{multline}
  \label{eq:chi_2D_final_form}
  \chi=\sqrt{\frac{P_{tot}}{P_{0}}}~~\text{with},~~~~~
  P_{0}=\frac{1}{2}\Re e \left[ \int_{\Omega} \left( \pmb{\phi}_{t}+ i\beta  \phi_{z}\mathbf{\hat{z}} \right) \times \right. ~~~~~~~~~~~~~~~~~~~~~~~~\\ \left. \left( \overline{  \frac{1}{i\omega \mu_{0}} \left[ \left( {\nabla}_{t} \times \pmb{\phi}_{t} \right)+ \left[ i\beta \left( \pmb{\nabla}_{t}  \phi_{z} - \pmb{\phi}_{t} \right) \right] \times ~\mathbf{\hat{z}}    \right]   } \right)  \cdot \mathbf{\hat{z}} ~~ \text{d}\Omega \right],	
\end{multline}
and $P_{0}$ is the integral of the current Poynting vector along $z$ direction. \eqref{eq:nl_perm_general} and~\eqref{eq:2D_functions_chi_Et_Ez}  (together with the weak formulation given in~\citep{zolla_foundations_2012}) form a general anisotropic nonlinear vector eigenvalue problem with $\beta$ being the complex eigenvalue and $\pmb{\phi}_{t}, \phi_{z}\mathbf{\hat{z}}$ are the complex eigenfunctions. To solve this nonlinear vector eigenvalue problem, we will use the fixed power algorithm, that uses a sequence of linear eigenvalue problems to compute the stationary nonlinear solutions following Algorithm~\ref{alg:fixed-power-2D-vectorial-case}.

\begin{algorithm}[H]
  \caption{Fixed power algorithm to solve the full vector nonlinear eigenvalue problem within the FEM approach for a nonlinear permittivity given by ~\eqref{eq:nl_perm_general} and~\eqref{eq:2D_functions_chi_Et_Ez}.}
  \label{alg:fixed-power-2D-vectorial-case}
  \begin{algorithmic}[1]
    \State We start with an initial guess for $\mathbf{E}(x,y)$ (usually the linear solution), which will be used in~\eqref{eq:nl_perm_general} to compute the nonlinear permittivity tensor $\tilde{\bar{\bar{\epsilon}}}_{t}$ and $\tilde{\bar{\bar{\epsilon}}}_{z}$. 
    \State  We use $\tilde{\bar{\bar{\epsilon}}}_{t}$ and $\tilde{\bar{\bar{\epsilon}}}_{z}$ to define the 2D vector linear  eigenvalue problem, then we compute $\pmb{\phi}_{t}$ and $\phi_{z}\mathbf{\hat{z}}$ with $\beta$ the corresponding complex eigenvalue.  These outputs are then used to compute the scaling factor $\chi$, for a given fixed value of the power $P_{tot}$ using~\eqref{eq:chi_2D_final_form}.
    \State The scaling factor $\chi$ will be used to compute the correct amplitude of the longitudinal and the transverse components of the electric field $\pmb{E}_{t}$ and $E_{z} \mathbf{\hat{z}}$, respectively using~\eqref{eq:2D_functions_chi_Et_Ez}. 
    \State The rescaled electric field will be used to update the nonlinear permittivity tensors $\tilde{\bar{\bar{\epsilon}}}_{t}$ and $\tilde{\bar{\bar{\epsilon}}}_{z}$  defined by ~\eqref{eq:nl_perm_general} which will be used as inputs for the next iteration.
    \State We repeat steps ($2$), ($3$), and ($4$) until the following criterion is satisfied: ${|\Re e(\beta^{i})-\Re e(\beta^{i-1})|}/{|\Re e(\beta^{i})|} < \delta, i \in [2,N_{max}],$
    where $\beta^{i}$ is the eigenvalue for the last step $i$ and $N_{max}$ is the maximum step number in the procedure. We set $\delta=10^{-6}$ such that in order to fulfil the criterion between 10 and 15 steps are needed depending on the waveguide parameters and the initial field used.
  \end{algorithmic}
\end{algorithm}
In this work, we will focus on isotropic structures for simplicity, however anisotropic configuration can be treated using our method. To start, we consider a symmetric 2D NPSW with an isotropic Kerr-type nonlinearity in the core (see the inset in Fig.~\ref{fig:structure2D}) such that the linear part of the core permittivity gives $\epsilon_{jj}=\epsilon_{l,core}$ and ${\alpha_{jj}}={\alpha}~~~\forall j \in {x,y,z}$ in which ${ {\alpha}}\approx  \epsilon_{0} c \Re e(\epsilon_{l,core}) n_{2} >0$~\citep{boyd_nonlinear_2008}, where $\epsilon_{0}$ is the vacuum permittivity and $n_{2}$ is the nonlinear refractive index of the core material. In this study, for the core, we consider $\epsilon_{l,core} = 3.46^2 + i \, 10^{-4}$ and $n_2= 2.\, 10^{-17}$m$^2$~/W corresponding to amorphous hydrogenated silicon~\citep{lacava_nonlinear_2013} at $\lambda=1.55~\mu$m; we consider gold for the linear metal claddings with permittivity $\epsilon_{clad}=-90+i10$, in addition, we use $\text{SiO}_{2}$ and air for the linear substrate and the linear cover regions, with  $\epsilon_{sub}=1.46^{2}$ and $\epsilon_{cov}=1.0$, respectively.
\begin{figure}[h]
\centering
\includegraphics[width=\columnwidth]{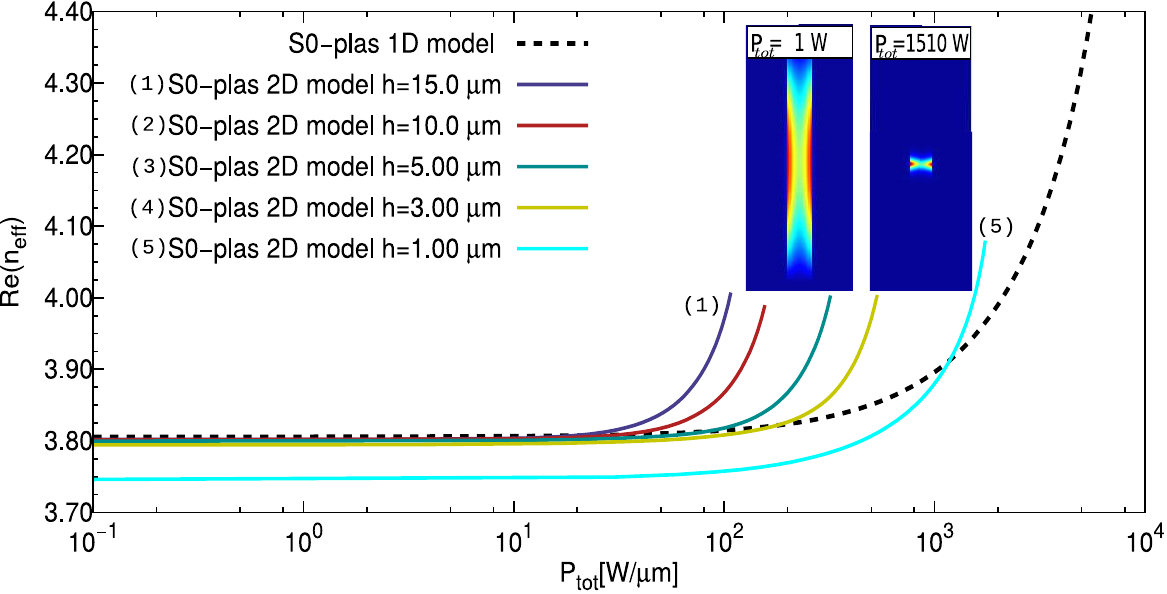}
\caption{Comparison between the nonlinear dispersion curve for the S0-mode obtained from the 1D model (dashed curve) and the  nonlinear dispersion curves for the S0-plas mode obtained from the 2D model with different height values $h$, $d_{core}=400$ nm, and $\text{W}_{\text{clad}}=0.5~\mu$m. The power for the 2D curves is normalized by the waveguide height $h$. The inset represents the field maps for S0-plas mode obtained from our 2D model with $h=15~\mu$m for two different power values.}
\label{fig:1D-to-2D-MIM}
\end{figure}

In Fig.~\ref{fig:structure2D}, we study the influence of the waveguide height $h$ on the real part of the effective index $\Re e(n_{eff})$  for the fundamental symmetric mode (denoted by S0-plas) in the linear regime. The horizontal dashed line represents the value of $\Re e(n_{eff})$ for S0-plas mode of the 1D model. As it can be seen in Fig.~\ref{fig:structure2D}, for waveguide heights above $4~\mu$m, $\Re e(n_{eff})$ obtained from the 2D model converges to the results obtained from the 1D model~\cite{walasik_plasmon-soliton_2016,elsawy_study_2017}. A similar convergence property is observed for the imaginary part of  $n_{eff}$. 

In Fig.~\ref{fig:1D-to-2D-MIM}, we show the importance of our 2D nonlinear modelling, by computing the nonlinear dispersion curves for the S0-plas mode of the 2D NPSWs (see the inset in Fig.~\ref{fig:structure2D}) with different height values $h$ (the power for those curves are normalized to the value of $h$ used). For the comparison, we also provide the results obtained from the 1D case (dashed curve). In Fig.~\ref{fig:1D-to-2D-MIM} as expected, in the low power regime, one notices that for 2D waveguides with small $h$, $\Re e(n_{eff})$ is smaller than the corresponding 1D value, as it can also be seen in Fig.~\ref{fig:structure2D}. At high power, the shape of the nonlinear dispersion curves obtained in the 2D case with  $h \gg \lambda$ is qualitatively similar to the one obtained in the 1D case (dashed curve). Nevertheless, the effective indices obtained in the 2D case change much more rapidly at high power than the one obtained in the 1D case, and that is due to the type of focusing and can be explained as follows.

At low powers, with $h \gg \lambda$ the 2D mode field profiles are broadened along the vertical direction ($y$-axis) and are  similar to the modes obtained in the 1D case (see the inset of Fig.~\ref{fig:1D-to-2D-MIM}), $\Re e(n_{eff})$ obtained from the 2D model converges to the one obtained from the 1D model. However, at high power, the mode profiles obtained in the 2D case are different from the ones obtained in the 1D case, since in the 2D case we achieve the focusing along both $x$ and $y$ directions (see the 2D field map at high power in the inset of Fig.~\ref{fig:1D-to-2D-MIM}), unlike in the 1D case where the focusing occurs along $x$ direction only.  These results underline the importance of our 2D nonlinear solver to provide an accurate  qualitative and quantitative description of the nonlinear effects in realistic 2D NPSWs distinct of those from 1D NPSWs~\cite{davoyan_nonlinear_2008,rukhlenko_exact_2011,salgueiro_complex_2014,walasik_plasmon-soliton_2016,elsawy_study_2017}. 
\begin{figure}[h]
\centering
\includegraphics[width=\columnwidth]{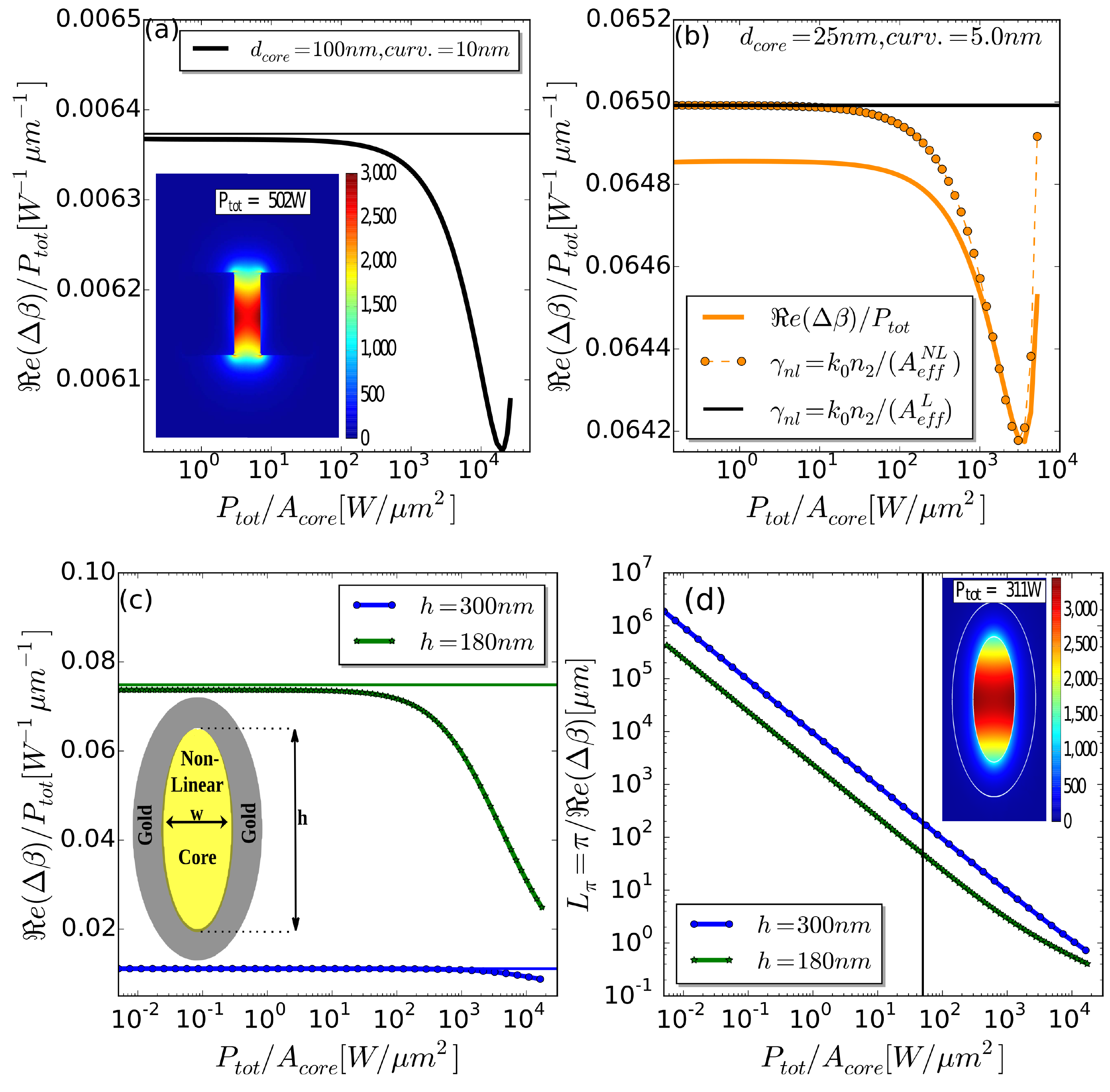}
\caption{(a-c) represent $\Re e(\Delta \beta)/P_{tot}$ as a function of the total power normalized by  the nonlinear core surface $P_{tot}/A_{core}$ for two different plasmonic waveguides. (a) and (b) for the 2D NPSW shown in the inset of Fig.~\ref{fig:structure2D} with  $h=250$ nm, $W_{clad}=200$ nm for two different core thicknesses; $d_{core}=100$ nm and $d_{core}=25$ nm, respectively. The inset in (a) represents $|\mathbf{E}_{t}|$ for the nonlinear S0-plas obtained for $d_{core}=100$ nm at $P_{tot}\approx502~W$. (c) and (d): for the nonlinear plasmonic nanoshell (see the inset in (c)) with height $h=300$ nm and $h=180$ nm, respectively (the core width is fixed to $w=100$ nm). (d) represents the waveguide length needed to obtain a $\pi$-phase shift, the insets in (d) represents $|\mathbf{E}_{t}|$ for the nanoshell with $h=180$ nm at $P_{tot}\approx311~W$.}
\label{fig:Gamma-nl-MIM-nanoshell}
\end{figure}

Now, we move to the nanophotonic NPSWs with $h \ll \lambda$ and $d_{core} \ll \lambda$. There is a fundamental issue when modelling such kind of configurations in the FEM even in the linear case. This issue is related to the numerical hot spot which are generated at the corners between the different materials; even for all-dielectric waveguides~\cite{andersen_field_1978}. The electric field is infinite at the corner between two different materials, moreover, its direction changes abruptly along its edges. The field singularity at the corners will not affect the global quantities like the effective index however, all the local quantities near to the singularity will be affected. 

In our case, we  treat the nonlinear problem by solving a sequence of linear eigenvalue problems  (see Algorithm~\ref{alg:fixed-power-2D-vectorial-case}), this means that we need to be sure that the field we use in each iteration does not have any singularity. Consequently, in this work, for all the nanophotonic NPSWs ($h \ll \lambda$ and $d_{core} \ll \lambda$) that we study below, all the sharp corners will be rounded with a small radius of curvature in order to avoid the singularity induced by the sharp corners~\cite{pitilakis_theoretical_2016}. For all the structures studied below, full convergence studies have been realized as a function of the mesh size (data not shown), in order to ensure the convergence of global and local quantities. 
Using our rigorous nonlinear solver, we now present general results, including at high powers, for the nonlinear parameter $\gamma_{nl}$, which is widely used in the literature to quantify the nonlinear waveguide strength~\cite{maslov_rigorous_2014,koos_nonlinear_2007,li_general_2017}. Actually, there are two ways to compute $\gamma_{nl}$; the first one is based on the effective mode area $A_{eff}$ of the linear field profile, such that using the nonlinear parameter $n_{2}$, one can write $\gamma_{nl}\approx{k_{0} n_{2}}/{A_{eff}}$~\cite{koos_nonlinear_2007,li_general_2017} where, $n_{core}$ is the linear core refractive index and $A_{eff}$ is the effective mode area as given in~\citep{koos_nonlinear_2007}. The second way of computing $\gamma_{nl}$ (eventually a complex number) is based on the change of the propagation constant induced by the power, such that, when the change is small, one gets~\cite{maslov_rigorous_2014}:
\begin{equation}
\label{eq:nl_param_gamm_nl_solver}
\gamma_{nl} (P_{tot}) = {\Delta \beta (P_{tot}) }/{P_{tot}}= {\beta(P_{tot})-\beta(P_{tot}=0)}/{P_{tot}}.
\end{equation}
\begin{figure}[b!h]
\centering
\includegraphics[width=\columnwidth]{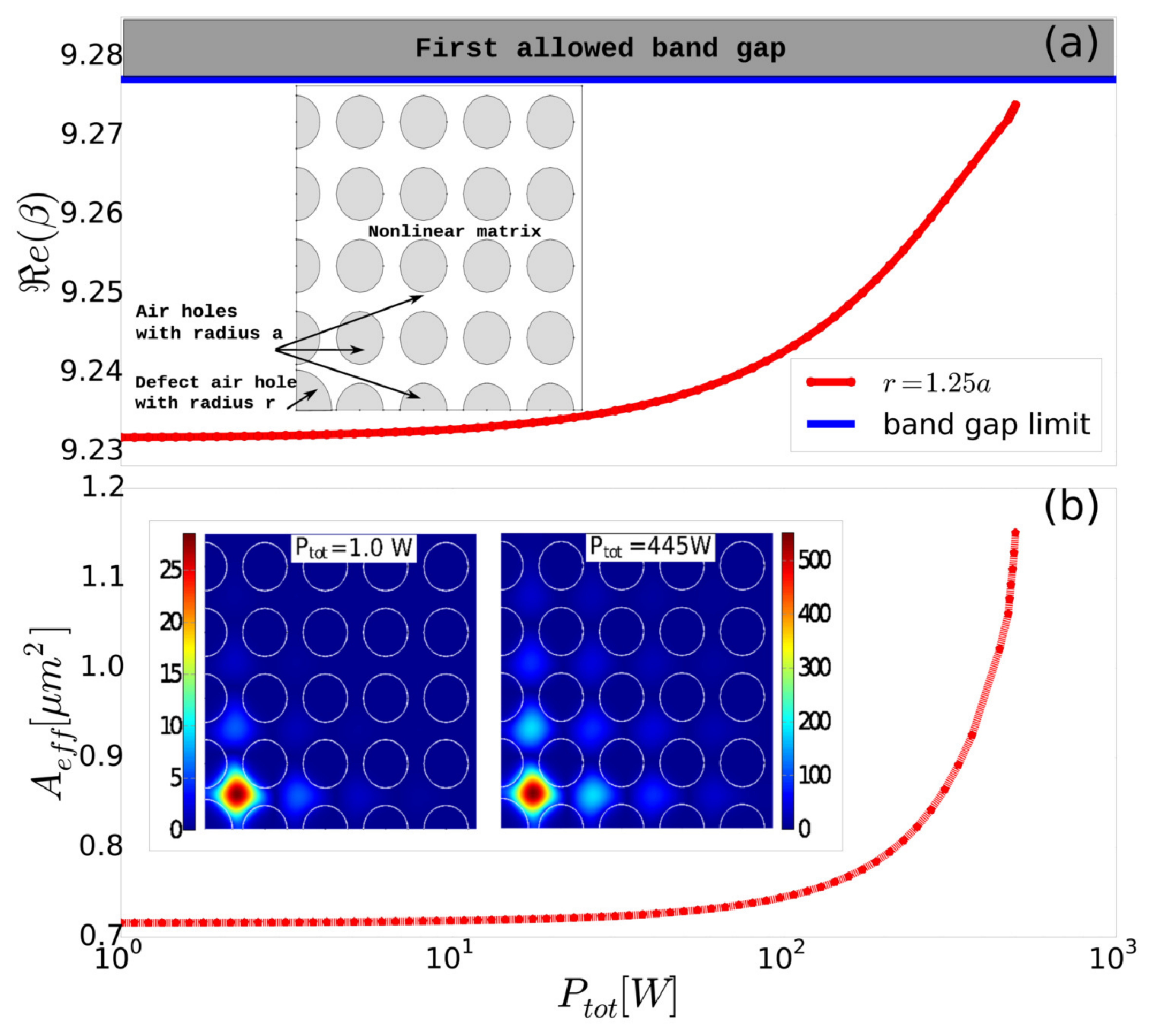}
\caption{Results as a function of $P_{tot}$ for the main nonlinear mode of a PCF with a core defect (see the inset in (a)) with $\epsilon_{l,core} = 2.47^2$, $n_2= 2.\, 10^{-17}$m$^2$~/W , and $\lambda=1.55 \,\mu$m. (a): $\Re e(\beta)$ and  (b): $A_{eff}$, its inset represents $|\mathbf{E}_{t}|$ at low and high power values. The pitch of the whole lattice is $\Lambda=0.95\lambda$, the radius of the lattice holes is $a=0.5\Lambda\times0.7375$, and the radius of the core defect hole is $r=1.25a$.}
\label{fig:PCF-accepteur-case}
\end{figure}
 In the low power  regime, it has already been shown that, in the lossless case, the conventional way of computing $\gamma_{nl}$~\citep{koos_nonlinear_2007} and~\eqref{eq:nl_param_gamm_nl_solver} coincide~\cite{maslov_rigorous_2014,sato_three-dimensional_2014}. Moreover, it was recently shown that  the definition given by~\eqref{eq:nl_param_gamm_nl_solver} is more general and valid even for high lossy waveguides, while the conventional definition $\gamma_{nl}\approx{k_{0} n_{2}}/{A_{eff}}$ is not valid for lossy waveguides~\cite{li_general_2017}. 
In Fig.~\ref{fig:Gamma-nl-MIM-nanoshell}, we go one step further by computing  $\gamma_{nl}(P_{tot})$ at high powers using \eqref{eq:nl_param_gamm_nl_solver} for two different 2D plasmonic waveguides made of the same materials. 
First, we study the 2D NPSW shown in the inset of Fig.~\ref{fig:1D-to-2D-MIM} for a fixed height $h=250$ nm and two different core thicknesses. The results in  Fig.~\ref{fig:Gamma-nl-MIM-nanoshell}(a-b) show that in the low power regime where $\Re e(\Delta \beta)/P_{tot}$ is constant, the two definitions of $\gamma_{nl}$ agree as expected. The small discrepancy between the two results shown in Fig.~\ref{fig:Gamma-nl-MIM-nanoshell} (b) is due to the losses, since the conventional definition of $\gamma_{nl}$ shown by the horizontal line is not fully valid for lossy waveguides as expected~\citep{li_general_2017}. However, at high powers where $\Re e(\Delta \beta)/P_{tot}$ is no longer constant (see Fig.~\ref{fig:Gamma-nl-MIM-nanoshell}), the conventional way of computing $\gamma_{nl}$ fails to predict the behaviour since it is based only on the linear field profiles. Using ${A^{NL}_{eff}}$, the effective mode area computed from the nonlinear mode profile in the definition of $\gamma_{nl}$ we obtain a similar behaviour as the one obtained from  $\Re e(\Delta \beta)$ (see Fig.~\ref{fig:Gamma-nl-MIM-nanoshell}(b)). The fact that $\gamma_{nl}(P_{tot})$ decreases and then increases is linked to its relation with $\Delta \beta/P_{tot}$:  the higher order terms in powers of $P_{tot}$ in the taylor expansion of  $\Delta \beta(P_{tot})$ must be considered when the power starts to be high enough. Second, to complete our results, in Fig.~\ref{fig:Gamma-nl-MIM-nanoshell} (c), we study $\Re e(\Delta \beta)/P_{tot}$ for a nonlinear plasmonic nanoshell~\citep{hossain_ultrahigh_2011} in which there is no corner avoiding field singularities. We show again the subtle behaviour of  $\Delta \beta $ for highly nonlinear plasmonic waveguides, which confirms the usefulness of our nonlinear solver in evaluating the nonlinear behaviour at low and high power levels. In Fig.~\ref{fig:Gamma-nl-MIM-nanoshell} (d), we study the waveguide length  needed to obtain a $\pi$-phase shift for the nanoshell configurations. In order to provide meaningful results, we take into account the damage power threshold for our nonlinear material ($5~GW/cm^{2}$) shown by a vertical line in Fig.~\ref{fig:Gamma-nl-MIM-nanoshell} (d).  Before the damage threshold, we can achieve a $\pi$-phase shift, using a small device length in the range of tens of micrometers while it requires millimeter long all-dielectric nonlinear slots~\citep{koos_nonlinear_2007,sato_three-dimensional_2014}.

The last example we treat with our vector nonlinear eigenmode solver is a nonlinear photonic crystal fiber. This kind of structure made of a subset of a periodic lattice of air holes embedded in a Kerr type nonlinear matrix with a fully solid defect core has already been studied in the scalar case~\citep{drouart_spatial_2008,ferrando_variational_2013}. In these works, the type of studied defect core corresponds to the limit case of a "donor" defect (no air hole at the core location): the main linear core localized defect mode comes from the first allowed band of the band diagram of the periodic structure and emerges in the semi-infinite forbidden band-gap~\citep{ferrando2000donor-acceptor,zolla_foundations_2012}. Here, we investigate a nonlinear fiber with a defect core of "acceptor" type (larger air hole at the core location)~\citep{ferrando2000donor-acceptor} in a square lattice of air holes, see inset of Fig.~\ref{fig:PCF-accepteur-case} (a). In this case, the  main linear core localized defect mode also comes from the first allowed band of the band diagram but it emerges in the first finite forbidden band-gap (see Fig.~4). 
The noteworthy behavior of this mode is found in its nonlinear regime when it moves towards the upper edge of this band-gap (blue line) . While the nonlinearity is of focusing type, the mode tends to delocalize in the lattice when the power is increased as shown in Fig.~\ref{fig:PCF-accepteur-case} (b) (field maps and effective area). This is a clear illustration of the interplay between the periodicity of the structure and the power controlled nonlinearity. To the best of our knowledge this is the first time such phenomenon is described using full vector Maxwell's equation.
In the scalar case,  the nonlinear behaviour of a comparable defect mode has already been studied but only at high power when the mode is located in the first semi-infinite band-gap but not in the first finite forbidden  band-gap~\citep{yang2003PCF}. In this last case,  with our solver we also find again that the modes focalizes with the power (data not shown)  as expected.

Thanks to our 2D vector eigenvalue nonlinear solver for Maxwell's equations, we have obtained three new results: the importance of the 2D model for NPSWs even at low power, a general interpretation of the link between $\gamma_{nl}$ and $\Delta \beta/P_{tot}$, and finally the illustration of the interplay between the  nonlinearity and the bad-gap edge giving defocusing effect from a focusing Kerr-term. These results confirm the need and the usefulness to pursue theoretical and numerical investigations of the full set of Maxwell's equations in both the low and high power nonlinear regimes.



\begin{thebibliography}{10}
\newcommand{\enquote}[1]{``#1''}

\bibitem{dekker2007ultrafast}
R.~Dekker, N.~Usechak, M.~F{\"o}rst, and A.~Driessen, \enquote{Ultrafast
  nonlinear all-optical processes in silicon-on-insulator waveguides,} Journal
  of physics D: applied physics \textbf{40}, R249 (2007).

\bibitem{kauranen_nonlinear_2012}
M.~Kauranen and A.~V. Zayats, \enquote{Nonlinear plasmonics,} Nature Photon.
  \textbf{6}, 737--748 (2012).

\bibitem{walasik_stationary_2014}
W.~Walasik, G.~Renversez, and Y.~V. Kartashov, \enquote{Stationary
  plasmon-soliton waves in metal-dielectric nonlinear planar structures:
  {Modeling} and properties,} Phys. Rev. A \textbf{89}, 023816 (2014).

\bibitem{davoyan_nonlinear_2008}
A.~R. Davoyan, I.~V. Shadrivov, and Y.~S. Kivshar, \enquote{Nonlinear plasmonic
  slot waveguides,} Opt. Express \textbf{16}, 21209 (2008).

\bibitem{rukhlenko_exact_2011}
I.~D. Rukhlenko, A.~Pannipitiya, M.~Premaratne, and G.~P. Agrawal,
  \enquote{Exact dispersion relation for nonlinear plasmonic waveguides,} Phys.
  Rev. B. \textbf{84}, 113409 (2011).

\bibitem{salgueiro_complex_2014}
J.~R. Salgueiro and Y.~S. Kivshar, \enquote{Complex modes in plasmonic
  nonlinear slot waveguides,} J. Opt. \textbf{16}, 114007 (2014).

\bibitem{walasik_plasmon-soliton_2016-1}
W.~Walasik, G.~Renversez, and F.~Ye, \enquote{Plasmon-soliton waves in planar
  slot waveguides. {II}. {Results} for stationary waves and stability
  analysis,} Phys. Rev. A \textbf{93}, 013826 (2016).

\bibitem{elsawy_study_2017}
M.~M.~R. Elsawy and G.~Renversez, \enquote{Study of plasmonic slot waveguides
  with a nonlinear metamaterial core: semi-analytical and numerical methods,}
  Journal of Optics \textbf{19}, 075001 (2017).

\bibitem{fujisawa_guided_2006}
T.~Fujisawa and M.~Koshiba, \enquote{Guided modes of nonlinear slot
  waveguides,} IEEE Photon. Tech. Lett. \textbf{18}, 1530--1532 (2006).

\bibitem{ferrando_spatial_2003}
A.~Ferrando, M.~Zacarés, P.~F.~d. Córdoba, D.~Binosi, and J.~A. Monsoriu,
  \enquote{Spatial soliton formation in photonic crystal fibers,} Opt. Express
  \textbf{11}, 452--459 (2003).

\bibitem{fujisawa2003PCF-NL}
T.~Fujisawa and M.~Koshiba, \enquote{Finite element characterization of
  chromatic dispersion in nonlinear holey fibers,} Opt. Express \textbf{11},
  1481--1489 (2003).

\bibitem{boyd_nonlinear_2008}
R.~W. Boyd, \emph{Nonlinear {Optics}} (Academic Press, 2008).

\bibitem{dular_general_1998}
P.~Dular, C.~Geuzaine, F.~Henrotte, and W.~Legros, \enquote{A general
  environment for the treatment of discrete problems and its application to the
  finite element method,} IEEE Trans. Magnet. \textbf{34}, 3395--3398 (1998).

\bibitem{drouart_spatial_2008}
F.~Drouart, G.~Renversez, A.~Nicolet, and C.~Geuzaine, \enquote{Spatial {Kerr}
  solitons in optical fibres of finite size cross section: beyond the {Townes}
  soliton,} J. Opt. A: Pure Appl. Opt. \textbf{10}, 125101 (2008).

\bibitem{rahman_numerical_1990}
B.~M.~A. Rahman, J.~R. Souza, and J.~B. Davies, \enquote{Numerical analysis of
  nonlinear bistable optical waveguides,} IEEE Photon. Tech. Lett. \textbf{2},
  265--267 (1990).

\bibitem{zolla_foundations_2012}
F.~Zolla, G.~Renversez, A.~Nicolet, B.~Kuhlmey, S.~Guenneau, D.~Felbacq,
  A.~Argyros, and S.~Leon-Saval, \emph{Foundations of {Photonic} {Crystal}
  {Fibres}} (Imperial College Press, 2012), 2nd ed.

\bibitem{lacava_nonlinear_2013}
C.~Lacava, P.~Minzioni, E.~Baldini, L.~Tartara, J.~M. Fedeli, and I.~Cristiani,
  \enquote{Nonlinear characterization of hydrogenated amorphous silicon
  waveguides and analysis of carrier dynamics,} Appl. Phys. Lett. \textbf{103},
  141103 (2013).

\bibitem{walasik_plasmon-soliton_2016}
W.~Walasik and G.~Renversez, \enquote{Plasmon-soliton waves in planar slot
  waveguides. {I}. {Modeling},} Phys. Rev. A \textbf{93}, 013825 (2016).

\bibitem{andersen_field_1978}
J.~Andersen and V.~Solodukhov, \enquote{Field behavior near a dielectric
  wedge,} IEEE Trans. Ant. Prop. \textbf{26}, 598--602 (1978).

\bibitem{pitilakis_theoretical_2016}
A.~Pitilakis, D.~Chatzidimitriou, and E.~E. Kriezis, \enquote{Theoretical and
  numerical modeling of linear and nonlinear propagation in graphene
  waveguides,} Optical and Quantum Electronics \textbf{48}, 243 (2016).

\bibitem{maslov_rigorous_2014}
A.~V. Maslov, \enquote{Rigorous calculation of the nonlinear {Kerr} coefficient
  for a waveguide using power-dependent dispersion modification,} Opt. Lett
  \textbf{39}, 4396--4399 (2014).

\bibitem{koos_nonlinear_2007}
C.~Koos, L.~Jacome, C.~Poulton, J.~Leuthold, and W.~Freude, \enquote{Nonlinear
  silicon-on-insulator waveguides for all-optical signal processing,} Opt.
  Express \textbf{15}, 5976--5990 (2007).

\bibitem{li_general_2017}
G.~Li, C.~M. de~Sterke, and S.~Palomba, \enquote{General analytic expression
  and numerical approach for the {Kerr} nonlinear coefficient of optical
  waveguides,} Opt. Lett \textbf{42}, 1329--1332 (2017).

\bibitem{sato_three-dimensional_2014}
T.~Sato, S.~Makino, Y.~Ishizaka, T.~Fujisawa, and K.~Saitoh,
  \enquote{Three-dimensional finite-element mode-solver for nonlinear periodic
  optical waveguides and its application to photonic crystal waveguides,} J.
  Lightwave Technol. \textbf{32}, 3409--3417 (2014).

\bibitem{hossain_ultrahigh_2011}
M.~M. Hossain, M.~D. Turner, and M.~Gu, \enquote{Ultrahigh nonlinear nanoshell
  plasmonic waveguide with total energy confinement,} Opt. Express \textbf{19},
  23800--23808 (2011).

\bibitem{ferrando_variational_2013}
A.~Ferrando, C.~Milián, and D.~V. Skryabin, \enquote{Variational theory of
  soliplasmon resonances,} J. Opt. Soc. Am. B. \textbf{30}, 2507 (2013).

\bibitem{ferrando2000donor-acceptor}
A.~Ferrando, E.~Silvestre, J.~Miret, P.~Andr{\'e}s, and M.~Andr{\'e}s,
  \enquote{Donor and acceptor guided modes in photonic crystal fibers,} Opt.
  Lett. \textbf{25}, 1328--1330 (2000).

\bibitem{yang2003PCF}
J.~Yang and Z.~H. Musslimani, \enquote{Fundamental and vortex solitons in a
  two-dimensional optical lattice,} Opt. Lett. \textbf{28}, 2094--2096 (2003).

\end{thebibliography}
\end{document}